# INTERBAND NODAL-REGION PAIRING AND THE ANTINODAL PSEUDOGAP IN HOLE DOPED CUPRATES

N. KRISTOFFEL AND P. RUBIN
*Institute of Physics, University of Tartu, Riia 142, 51014 Tartu, Estonia*

Abstract- Recent experimental findings show that the pairing interaction in hole-doped cuprates resides in the nodal (FS arcs) region accompanied by the separate antinodal pseudogap. A corresponding multiband model of cuprate superconductivity is developed. It is based on the electronic spectrum evolving with doping and extends authors' earlier approaches. The leading pair-transfer interaction is supposed between the itinerant (mainly oxygen) band and a nodal defect (polaron) band created by doping. These components are overlapping. The defect subband created in the antinodal region of the momentum space does not participate in the pairing. A supposed bare gap separating it from the itinerant band top disappears with extended doping. The corresponding antinodal pseudogap appears as a perturbative band structure effect. The low energy excitation spectrum treated in the mean-field approximation includes two nodal superconducting gaps and one pseudogap. The behaviour of these gaps and of other pairing characteristics agree qualitatively with the observations on the whole doping scale.

Keywords: cuprates, multiband superconductivity, phase diagram, gaps, doping

## 1. Introduction

Cuprate superconductivity mechanism remains elusive. A huge package of information on cuprate low-energy excitations [1,2] has been collected by diverse experiments. However, numerous improvements, corrections and precesions have been necessary to get comparable data for distinct conditions. Some essential results have been obtained quite recently, e.g. the coexistence of the pseudogap (PG) and of the superconducting gap (SCG) with the own energy scales at fixed doping [3-10]. The extraction of a true SCG from the background becomes possible [11,12]. The presence of the Fermi surface (FS) fragments in the form of arcs at underdoping [5,8,10,13] has been controlled by measuring the Shubnikov-de Haas oscillations [14-16]. The distinct reveal of the two SCG in cuprates succeed in [6] which was expected for long.

Recently one converged to the knowledge that at underdoping the energy scales of the large PG and of the smaller SCG are different by genealogy [4,5,10]. These gaps develop from different momentum space regions: the PG is an antinodal, and the SCG a nodal event [3,4,9,17,18]. The PG is extrinsic with respect of the superconductivity, i.e. different mechanisms in the superconductivity playground $CuO_2$ planes seem to be hidden behind these phenomena [18-20].

Contrary to the PG the superconducting pairing develops mostly on the FS arc segments around the nodal $\left(\frac{\pi}{2},\frac{\pi}{2}\right)$-type Brillouin zone positions [6,8,9,12,13,21,22]. The formation and expansion of these segments is due by doping. The introduced carriers concentrate first here [5]. The superconductivity created on arcs is accompanied by a sharp spectral coherence peak corresponding to the "arc metal" quasiparticles (QP) [21]. Extended doping brings wider momentum space regions into the game and a common FS will be built up.

The SCG rises and the PG diminishes with doping before the maximal $T_c$ is reached. In the underdoped regime the energy scales of these gaps are remarkably different. The PG disappears with overdoping. At this it does not transform into a SCG with entering the superconductivity dome [3]. The QP-s corresponding to these two type of gaps coexist [10] below the superconducting transition temperature ($T_c$). The PG QP-s remain here non-coherent. The PG is not very material sensitive, the SCG is [13]. The PG fills in with thermal excitation and fits a temperature driven metal-to-insulator transition under the doping concentration where it vanishes [23,24].

One has usually classified cuprates as two-gap systems having in mind the presence at least of one SCG and one PG. At present a long list of real two- (or multi-) gap superconductors which show multiple SCG-s is known. Cuprates seem to enter this community also. In fact, numerous direct and indirect experimental and theoretical results crowned by the observation of two SCG-s [6] positione the cuprates into the class of multigap superconductors with a multiband superconductivity mechanism. We refer here to the reviews [25-28] and some recent theoretical approaches [29-37]. Authors multiband approaches to cuprate superconductivity are based on the nonrigid nature of the electron spectrum reorganized by the necessary doping [35-40]. Interband pairing channels are opened on this background including the appeared defect-polaron type new bands. The present work takes into account novel findings mentioned, especially the exposed "arc superconductivity". The corresponding model includes only one antinodal nonmagnetic PG besides two SCG-s. In the present model only one nonmagnetic PG appears, cf. [35-37].

## 2. The model

Strong electronic correlations cause nonrigid behaviour of cuprate electron spectrum under doping. Perturbed distribution of states can include new band components and spectral gaps. Besides the itinerant carriers defect-polaron type particles have been identified in doped cuprates. A corresponding two-component scenario of cuprate superconductivity has been formulated [41-43]. On these two sorts of carriers a phase separation in the material becomes possible [44]. Sample regions rich of doped carriers create new defect-polaron type bands, the regions poor of them support the weakened itinerant band(s). As the result, the energetic spectrum near the FS is essentially changed and acquires multiband nature [45-49]. In the case of hole doping the actual FS evolution embraces the region near the top of the itinerant (mainly oxygen) band. Electron doping comprises the whole parent Hubbard antiferromagnetic spectrum. Defect bands are then induced under the top of the UHB.

The present work uses a modified model (cf. [37]) of hole doped cuprate defect subbands to follow more closely the novel findings mentioned in the Introduction. We use largely the foundation arguments and ideas from our earlier approaches [35-40] in coming to the model illustrated in Fig.1. The itinerant valence band ($\gamma$) with nodal top (energy zero) symmetry overlaps the nodal defect $\beta$-band states. The latter occupy energies from zero to ($-\beta c^2$). At this c is a measure of doped hole concentration. The itinerant band states of weight (1-c) end at (-D). The loosed weight charges equally by c/2 the defect $\beta$- and $\alpha$-bands. The latter evolves in antinodal region between d and $d-\alpha c^2$. The 2D densities of states ($CuO_2$ plane) of defect subbands ($\rho_{\alpha,\beta}$) $1/(2\alpha c)$ and $1/(2\beta c)$ depend on doping; $\rho_\gamma = (1-c)\ D^{-1}$. The bare $\beta$-$\gamma$ overlap and the $\rho_\beta(c)$ dependence take account of the residing "arc superconductivity".

The supposed bare gap between the $\alpha$-band bottom and the top of the itinerant band (ZRS binding energy?) closes at $c_\alpha = (d/\alpha)^{1/2}$. This gap acts as an extrinsic source for the antinodal PG as the result of the perturbation excerted by doping.

A further original moment consists in the appearance of the interband pairing channel opened on this spectrum. It consists in the interband pair-transfer interaction [26] between the overlapping $\beta$- and $\gamma$-bands. The corresponding characteristic constant $W = W_{\beta-\gamma}$ can include Coulombic and electron-phonon (interband!) contributions [26]. The analogous $\alpha$-$\gamma$ antinodal-nodal pairing is ignored by symmetry arguments, cf. [35] and experimental appointments, especially by the symmetry of the pseudogap.

There are three specific dispositions of the chemical potential ($\mu$) and the band components. At $c < c_\alpha$, $\mu$ intersects the overlapping $\beta$-$\gamma$ bands as

$$\mu_1 = -\beta c^2 \left[ 1 + (1-c)\frac{2\beta c}{D} \right]^{-1} . \qquad (1)$$

At $c>c_\alpha$, all the bands overlap, the α-band bottom reaches $\mu_1$ at $c_1$ $(\mu_1 = d - \alpha c_1^2)$ and for larger dopings

$$\mu_2 = -\beta(2\alpha c^2 - d)\left[ \alpha + \beta + (1-c)\frac{2\alpha\beta c}{D} \right]^{-1} . \qquad (2)$$

This regime can end when the situation $\mu_2 (c_2) = -\beta\, c^2_2$ is reached where the chemical potential leaves the β-band, and is given by

$$\mu_3 = (d - \alpha c^2)\left[ 1 + (1-c)\frac{2\alpha c}{D} \right]^{-1} . \qquad (3)$$

Now the effective β-γ pairing disappears as also the resonance of corresponding FS -s.

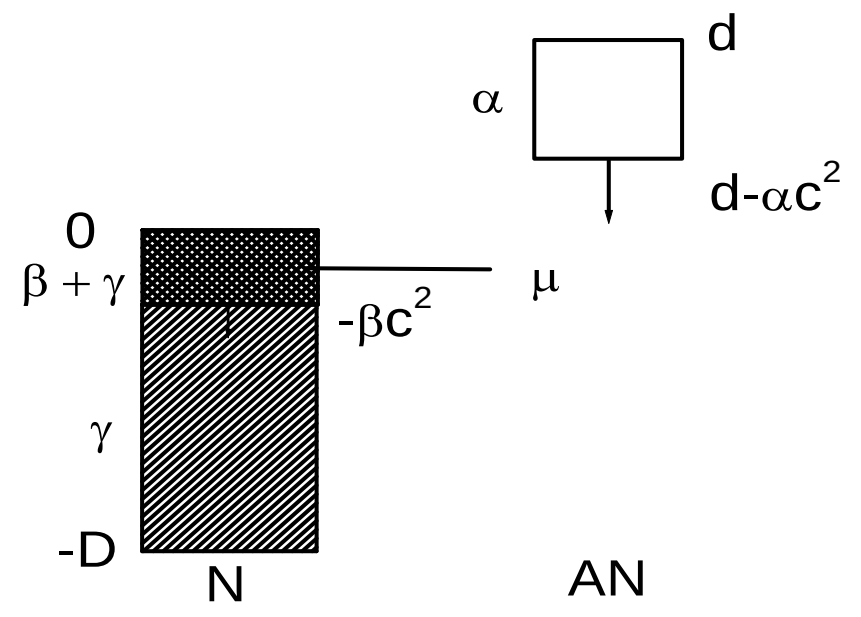

Fig. 1. The proposed energetic scheme for a hole-doped cuprate. In the nodal region there are the itinerant band γ (0>E>-D) and the defect band β (0>E>-$\beta c^2$). The antinodal defect band (α) occupies states between d and d-$\alpha c^2$. Note the defect bands bottom descending with doping.

The composite trend of $\mu$ shown in Fig. 2 follows the data for hole doped cuprates [49,50]. The nearly quadratic behaviour of $\mu$ with doping has stimulated the choice far the defect bands bottom lowering.

## 3. Low-energy excitations

The calculations on the model have been made using the mean-field version of the pair transfer mediated superconductor Hamiltonian

$$H=\sum_{\sigma,\vec{k},s}\varepsilon_\sigma(\vec{k})a^+_{\sigma\vec{k}s}a_{\sigma\vec{k}s}+\Delta_\gamma\sum_{\vec{k}}\left[a_{\gamma\vec{k}\uparrow}a_{\gamma-\vec{k}\downarrow}+a^+_{\gamma-\vec{k}\downarrow}a^+_{\gamma\vec{k}\uparrow}\right]-\Delta_\beta\sum_{\vec{k}}\left[a_{\beta\vec{k}\uparrow}a_{\beta-\vec{k}\downarrow}+a^+_{\beta-\vec{k}\downarrow}a^+_{\beta\vec{k}\uparrow}\right], \quad (4)$$

with the SCG defined as

$$\begin{cases}\Delta_\gamma=2W\sum_{\vec{k}}\left\langle a_{\beta\vec{k}\uparrow}a_{\beta-\vec{k}\downarrow}\right\rangle\\ \Delta_\beta=2W\sum_{\vec{k}}\left\langle a_{\gamma-\vec{k}\downarrow}a_{\gamma\vec{k}\uparrow}\right\rangle\end{cases}. \quad (5)$$

Usual designations are used with band energies counted from the chemical potential $\varepsilon_\sigma=\xi_\sigma-\mu$. The coupled gap system looks as ($\theta = k_BT$)

$$\begin{cases}\Delta_\gamma=W\Delta_\beta\sum_{\vec{k}}E_\beta^{-1}(\vec{k})th\dfrac{E_\beta(\vec{k})}{2\Theta}\\ \Delta_\beta=W\Delta_\gamma\sum_{\vec{k}}E_\gamma^{-1}(\vec{k})th\dfrac{E_\gamma(\vec{k})}{2\Theta}\end{cases} \quad (6)$$

with the Bogolyubov QP energies as $E_\sigma(\vec{k})=\pm\sqrt{\varepsilon_\sigma^2(\vec{k})+\Delta_\sigma^2}$.

The supercarrier density is given by

$$n_s=\frac{1}{2}\left[\sum_{\vec{k}}\frac{\Delta_\gamma^2\left(\vec{k}\right)}{E_\gamma^2\left(\vec{k}\right)}th^2\frac{E_\gamma(\vec{k})}{2\Theta}+\sum_{\vec{k}}\frac{\Delta_\beta^2\left(\vec{k}\right)}{E_\beta^2\left(\vec{k}\right)}th^2\frac{E_\beta(\vec{k})}{2\Theta}\right]. \quad (7)$$

Low-energy QP excitations of the present model are represented by two SCG-s $\Delta_\beta$ and $\Delta_\gamma$, and one PG $\Delta_{p\alpha}$. The SCG $\Delta_{\beta,\gamma}$ are of defect and itinerant genealogy correspondingly. The detailed symmetry of them does not follows from the model. In the case of two s-wave order parameters the gaps are of opposite signs at W>0 and the larger one corresponds to the component with smaller density of states [26]. Pure d and s, or mixed (d-s) ordering symmetries are allowed in two-band models according to the doping level and temperature. Extreme dopings favour separated d and s superconducting gap symmetries, cf [51,52].

The antinodal PG excitation energies lie in the interval between the α-band bottom and the chemical potential lying in the β-γ overlapping region, so that $\Delta_{p\alpha} = |\xi_\alpha - \mu_1|_{\min} = |d - \alpha c^2 - \mu_1|$. This PG vanishes at $c_1$, where the α-band starts to participate in determining of the chemical potential inside of all the three overlapping α, β, γ bands. The α PG enters the energy region occupied at $T<T_c$ by the nodal SCG-s. In this manner the nodal superconductivity coexists with incoherent antinodal PG excitations. Note that in the case $W_{\alpha\text{-}\gamma} \neq 0$ with the antinodal $\Delta_\alpha$ SCG present, $\Delta_{p\alpha}$ will be at $T<T_c$ and $c<c_1$ represented by $\left[(\xi_\alpha - \mu)^2_{\min} + \Delta^2_\alpha\right]^{1/2}$, and at $c>c_1$ by $\Delta_\alpha$ [37,38].

Extrinsic models of the PG attribute it often to the formation of an ordering of different nature [19,52-58] which competes with the superconductivity by exploiting a part of density of states necessary for the pairing. In our model the SCG-s and the PG does not compete as ordering characteristics. The gapped α-band belongs to the superconductivity background prepared by doping in which specific distortive lattice structures can appear.

## 4. Some results

The illustrative calculations have been made using the following parameters: D=2; d=0.3 [48,53], α=1; β=0.3 and W=0.28 (eV). The hole doping concentration has been scaled by the choice p=0.4 c, which exposes the maximal $T_{cm}$ = 146 K at p =0.16. One has $p_\alpha$=0.22; $p_1$=0.26; $p_2$=0.27.

The calculated superconductivity ($T_c$) dome and the antinodal PG are shown in Fig. 2. This is roughly an overall accepted image. The dome form driven by doping is typical for the interband pairing on overlapping bands with traveling chemical potential. It appeared in earliest multiband approaches to the cuprate superconductivity [26]. The spin-glass region with p<0.05 remains out of the scope  of the present work.

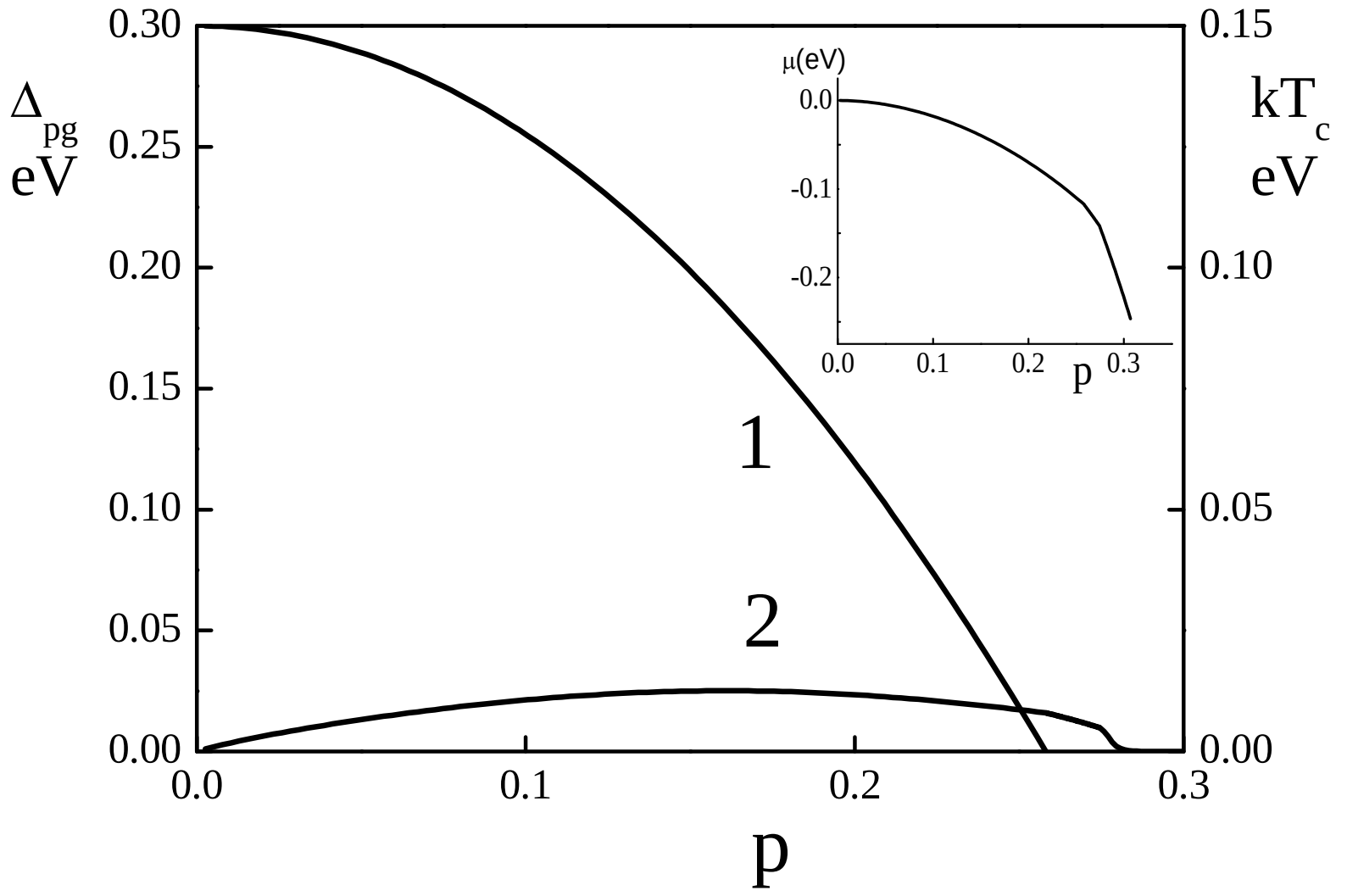

Fig. 2. The calculated antinodal pseudogap (1) and the $T_c$ dome (2) on the doping scale. The insert shows the behaviour of the chemical potential.

Various routes of the vanishing of the antinodal PG in the overdoped region have been discussed [18]. In our model the PG is not a measure of the pairing strength. $\Delta_{p\alpha}$ does not transform on the energy scale into $T_c$ or a SCG on the passage down towards zero inside the superconductivity domain, cf.[56]. In the normal state one must be able to follow its way to zero. In the superconducting state this can be masked in some extent by the residing SG gaps in this energy region. Presumably the different spectral windows (node, antinode) can help here for the differentiation. However, in reality these regions are spread in the momentum space as also the effects transmitted by them.

The vanishing of the PG influences $\mu$, but the metallization of the $\alpha$-particles behind $p_1$ does not enhance $T_c$ because $W_{\alpha\text{-}\gamma}=0$. Below $p_1$ the nodal particles are metallic and the antinodal ones remain insulating. The detailed route of $\alpha$ PG to zero depends on the properties of doping created defect (midgap) bands with the associated variety of lattice reconstructions (stripes etc). In the case where $W_{\alpha\text{-}\gamma}\neq 0$, as the calculations show, the vanishing of the PG is accompanied by a splash in $T_c$ (until the formation of a dome with two maxima). At comparable footing of both pairing channels on all overlapping band components resonating with $\mu$ an expressed $T_c$ maximum is built up [35,37]. In this sense the ending of a PG near $T_{cm}$ can point on the remarkable contribution of the antinodal part of the spectrum to pairing (then behind $p_1$ the normal state antinodal spectrum will be empty). The journey of the $\alpha$ PG on the doping scale is directly connected with our original description with moving defect band bottoms. It can be ascribed to diminishing perturbative action of carriers adding to the defective collective.

From the point of view of the present model there will be essential to control experimentally the participation of two “band” components (or SCG) in the arc superconductivity. In what extent it will be possible for the overlapping constituents of mainly same symmetry may be a question. In the LaSrCuO-system one has found in this region a second (smaller) pseudogap [59]. This can point to the presence of a subsystem borning as gapped from the itinerant band, cf. [37]. Note also the presence of the second $T_c$-maximum in this region (the 1/8 anomaly).

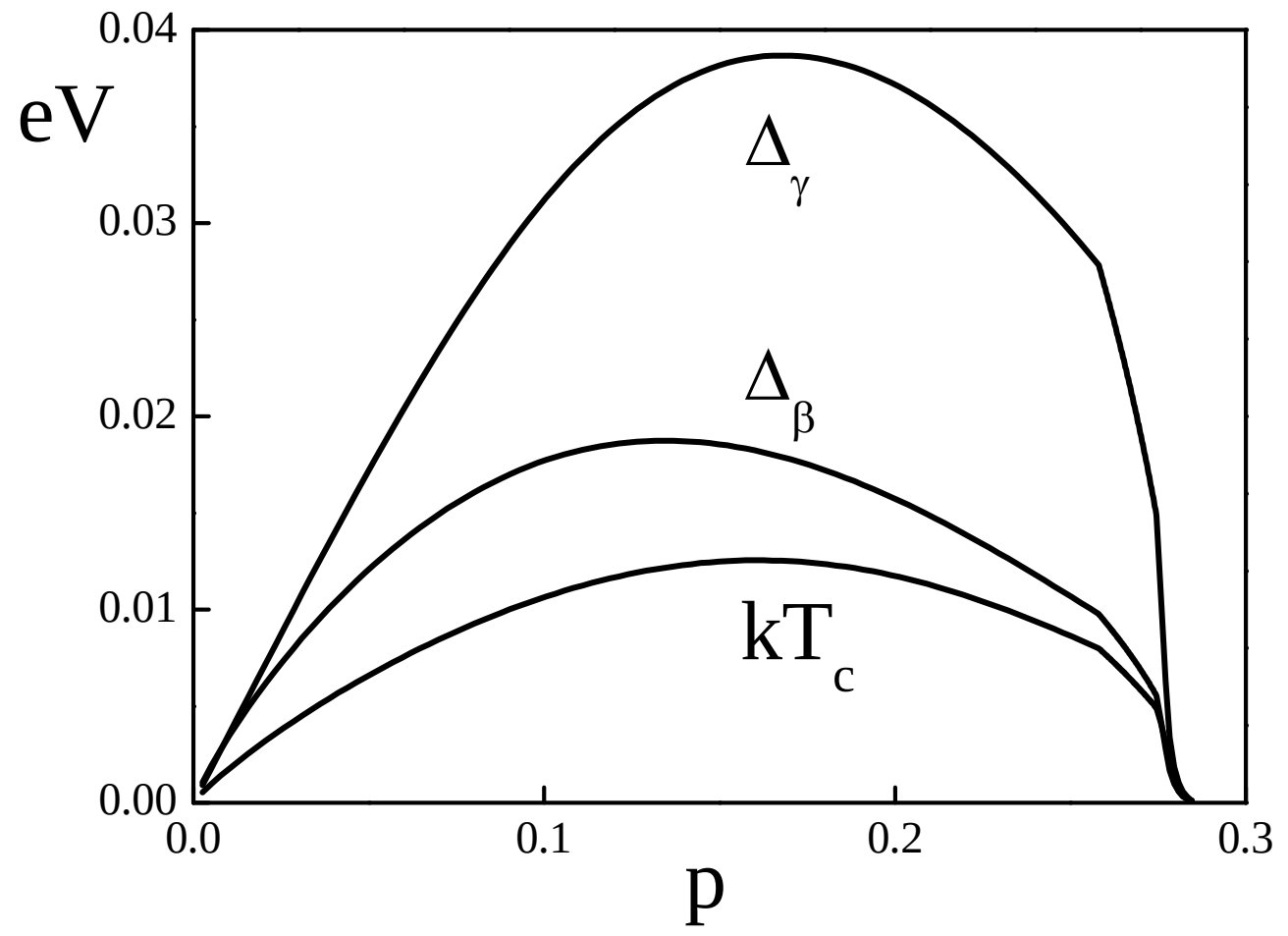

Fig. 3. The zero temperature superconducting gaps and the transition temperature dependences on doping.

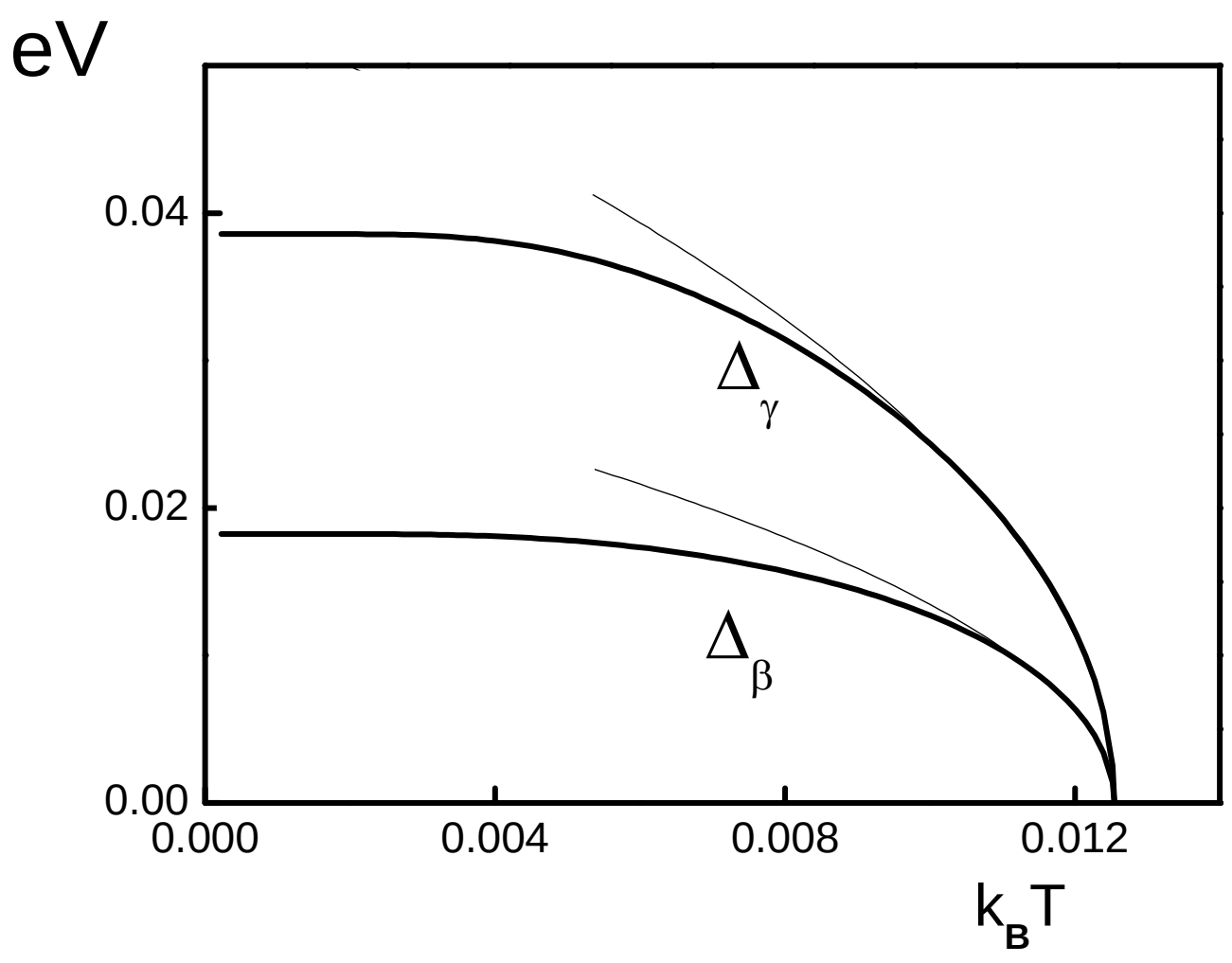

Fig. 4. The dependence of the superconducting gaps (p=0.16) on temperature. The thin lines control the $(1\text{-}T/T_c)^{1/2}$ law.

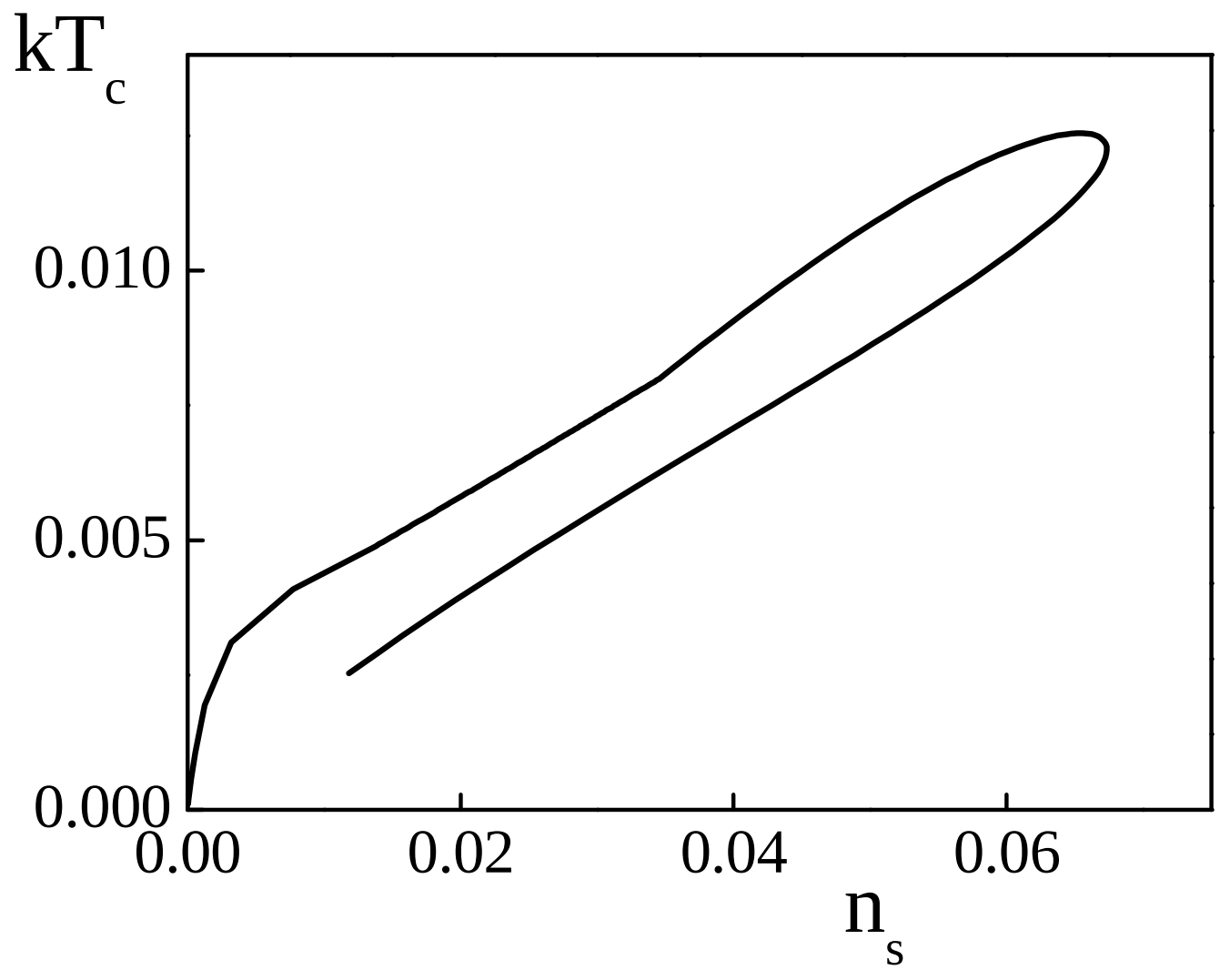

Fig. 5. The Uemura-plot representing the relation of $k_BT_c$ and $n_s$ on rising and decreasing arm of $T_c(p)$.

The zero-temperature nodal SCG-s are given in Fig. 3 and show also the typical bell-like behaviour on doping representing the grow and decrease of the supercarrier density ($n_s$). The dependence of these gaps on temperature is illustrated in Fig. 4. The supercarrier density does not equal to the doped carrier density and its common development with the transition temperature can be traced by the "Uemura-plot" in Fig. 5. The BCS universality is broken by the opposite dependences and magnitudes of $2\Delta_{\beta,\gamma}/kT_c$ on doping.
The results of the present work are in qualitative agreement with experimental findings. The presented model delivers in the natural way the widely accepted energetic phase diagram of a representative hole-doped cuprate superconductor. The multiband nature of the cuprate superconductivity by interband pairing finds support. The proposed plausible material background needs verification by *ab initio* calculations.

This work was supported by the Estonian Science Foundation grant No 7296.